\documentclass[twocolumn]{book}
\usepackage[dvips]{graphicx,color}
\usepackage{makeidx,universe}

\makeauthorindex

\BookTitle{Proceedings of the XXIX PHYSICS IN COLLISION}

\CopyRight{\copyright 2009 by Universal Academy Press, Inc.}

\begin{document} 

\pagenumbering{arabic}

\chapter{%
{\LARGE \sf
Review of Top Quark Measurements } \\
{\normalsize \bf 
Bernd Stelzer,} {\normalsize on behalf of the CDF and D0 Collaborations } \\
{\small \it \vspace{-.5\baselineskip}
Simon Fraser University, 8888 University Dr., Vancouver, BC, V5A 1S6, Canada
}
}

  \baselineskip=10pt 
  \parindent=10pt    

\section*{Abstract} 
Fermilab's Tevatron accelerator is recently performing at record luminosities that
enables a program systematically addressing the physics of top quarks.
The CDF and D0 collaborations have analyzed up to 5 fb$^{-1}$ of proton 
anti-proton collisions from the Tevatron at a center of mass energy of 1.96 TeV. The
large datasets available allow to push top quark measurements to higher and higher precision
and enabled the recent observation of electroweak single top quark production at the Tevatron.
This article reviews recent results on top quark physics at the Tevatron and provides a 
brief outlook on early top quark measurements at the Large Hadron Collider at CERN,
scheduled to restart in November 2009.

\section{Introduction}
The discovery of the top quark by the CDF and D0 collaborations at Fermilab in 1995 \cite{cdfd0topdisc} 
ended a 20 year quest for the top quark and completed the three-generation structure of the standard model (SM).
This observation marks the beginning of top quark physics, a new rich field in particle physics. 
The top quark is distinguished by its large mass, close to the scale of electroweak symmetry breaking
$v\sim$~246 GeV, while all other observed fermions have masses that are a tiny fraction of this energy. 
The origin of this unique property remains mostly a mystery, except that the top quark couples strongly to 
the mechanism of electroweak symmetry breaking (EWSB). It is speculated whether the top quark 
may play a fundamental role in EWSB or is special otherwise.
This makes top quark physics an attractive topic to look for physics beyond the SM.
New physics could allow the top quark to participate in new dynamics which could modify top quark couplings
and alter top quark production and decay. 
Studying the top quark in detail may provide hints of new physics that affect top quark properties.
Moreover, the top quark sample could in fact be 'contaminated' by exotic sources of top quarks such as SUSY or new resonances decaying to top quark pairs. The Tevatron top physics program explores all these possibilities. 

The large mass of the top quark leads to some interesting features. The top quark decays almost 
exclusively through the single mode $t\rightarrow Wb$. The on-shell $W$-boson can then 
decay leptonically or hadronically with coupling strengths given by the  Cabibbo-Kobayashi-Maskawa (CKM) matrix.
The top quark decay proceeds extremely fast, in less than 10$^{-24}$s, which is shorter than the time scale to form hadrons.  
Hence, the top quark provides us with the unique opportunity to study a bare quark.

\section{Production of Top Quarks} 
Until the Large Hadron Collider (LHC) turns on later this year, top quarks are exclusively produced at Fermilab's Tevatron 1.96 TeV proton anti-proton collider. Top quark production is a rare process at the Tevatron. Only one in 10 billion inelastic collisions features top quarks while the rate for generic QCD jet production is many orders of magnitude larger.

Within the SM, top quarks are predominantly produced in pairs via the strong interaction and singly via the electroweak interaction. 

\subsection{Top Quark Pair Production} 
\begin{figure}[h]
\begin{center}
  \includegraphics[width=.40\textwidth]{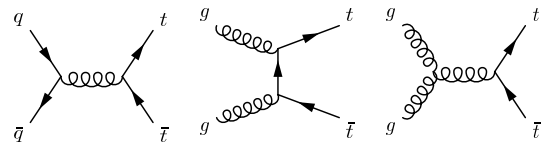}
   \caption{Top quark pair production at leading order through quark annihilation (left) and gluon fusion (middle, right).}
 \label{fig:cdfttprod}
\end{center}
\end{figure}

The main source of top quarks at the Tevatron is top quark pair production via the strong interaction as shown in Figure \ref{fig:cdfttprod}
At the Tevatron, production through quark annihilation contributes to about 85\% of the total rate, while gluon fusion process amounts to about 15\%. At the LHC, the contributions are expected to be roughly reversed.
The theoretical cross section has been calculated at Next to Leading Order (NLO) to be $\sigma_{t\bar{t}}=7.4^{+0.5}_{-0.7} $~pb at the Tevatron, assuming
a top quark mass of 172.5 GeV \cite{ttbarTheory}.
\begin{figure}[b]
\begin{center}
  \includegraphics[width=.40\textwidth]{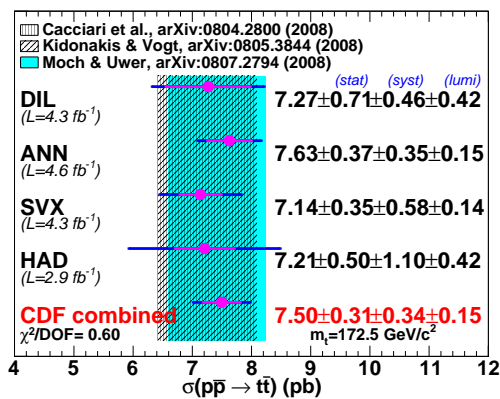}
   \caption{Summary of CDF top pair production cross section measurements, assuming a top quark mass of 172.5 GeV.}
 \label{fig:cdfttxsecs}
\end{center}
\end{figure}
The presence of two top quarks gives rise to unique event signatures classified according to the decay mode of the $W$ bosons. The most important (or golden) channel is the lepton + jets channel where one $W$ boson decays leptonically into a charged lepton ($e,\mu$) and a neutrino, while 
the second $W$ boson decays into two jets. This channel features a large branching ratio of about 34\% with manageable backgrounds, and allows for the full reconstruction of the event kinematics. The channel with the highest signal purity is the dilepton channel. Both $W$ bosons decay to a lepton ($e,\mu$) and a neutrino. The branching ratio for dilepton events is about 6\%. More challenging channels, due to large backgrounds, are the all-hadronic channel with a branching fraction of about 46\% and events with hadronic tau decays that contribute with a branching ratio of about 14\% in tau + jets and tau + lepton signatures.
\begin{figure}[b]
\begin{center}
\includegraphics[width=.40\textwidth]{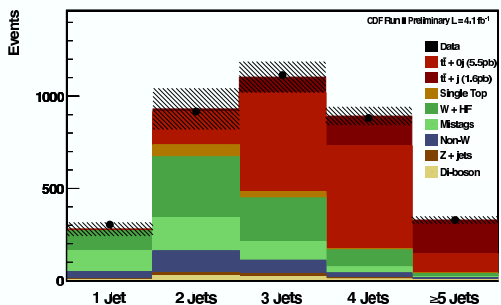}
\caption{Measurement of top pair production cross section in the lepton + jets channel by CDF. The plot shows data
as well as signal and background simulation for events with two or more $b$-tagged jets.}
\end{center}
\label{fig:cdfttxsec}
\end{figure}

Measuring the $t\bar{t}$ production cross section is important because a significant difference between the measurement
and the theoretical prediction indicates the presence of new physics. 
Depending on the final state signature, the measurement requires a detailed understanding of backgrounds and
good identification of leptons, jets, missing transverse energy, and efficient $b$-quark tagging.
To maximize sensitivity to a possible new physics signal, both, CDF and D0 have measured the production cross section in all
final state signatures. CDF has recently updated four measurements using the largest available datasets, close to 5 fb$^{-1}$.
One measurement was performed in the dilepton channel, one measurement in the all-hadronic channel and two measurements 
in the lepton+jets channel. The two measurements in the lepton+jets channel employ different strategies for
signal extraction. One uses an artificial neural network based on seven kinematic and topological observables.
The other analysis uses displaced secondary vertex $b$-tagging to purify the sample as shown in Figure \ref{fig:cdfttxsec}
Both measurements are systematically limited. The uncertainty on luminosity (6\%) has previously been the leading systematic.
Both analyses measure the ratio of the top pair cross section to the cross section of $Z$ boson production and
normalize to the theoretical $Z$ cross section which makes the analyses nearly free of the luminosity uncertainty.

The new CDF measurements are summarized in Figure \ref{fig:cdfttxsecs}
The results indicate consistency among final states and consistency with the theoretical prediction.
The combination of all measurements yields $\sigma_{t\bar{t}} = 7.5 \pm 0.31 (stat) \pm 0.34 (syst) \pm 0.15 (Z_{theory})$~pb
assuming a top quark mass of 172.5 GeV/c$^2$.
A similar combination performed by the D0 collaboration, including tau final states, yields. $\sigma_{t\bar{t}} = 8.18^{+0.98}_{-0.87}$~pb, assuming a top quark mass of 170 GeV/c$^2$ \cite{d0TTcomb}.
The precision on the CDF measurement has reached 6.5~\% which is now smaller than the uncertainty in the 
theoretical prediction. Therefore, improvements in the theoretical calculations will be required to
sustain sensitivity to new physics using these measurements.

\subsection{Single Top Quark Production} 
\begin{figure}[t]
\includegraphics[width=1.1\columnwidth]{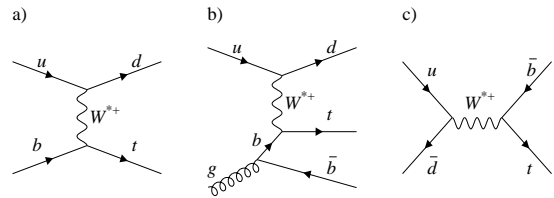}
\vskip -1.1cm
\caption{Representative Feynman diagrams of single top quark production. Figures (a) and (b)
are $t$-channel processes, and figure (c) is the $s$-channel process.}
\label{fig:sttfeyn}
\end{figure}

Top quarks are also expected to be produced in electroweak interactions, one at a time,
through a $s$-channel or $t$-channel exchange of a virtual $W$ boson, as
shown in Figure ~\ref{fig:sttfeyn} The cross sections for these interactions 
have been calculated at NLO $\sigma_{t}=2.05\pm 0.22$~pb~ for the $t$-channel and $\sigma_{s}=0.95\pm 0.08$~pb~
for the $s$-channel \cite{stxs}, respectively, assuming a top quark mass of 172.5 GeV/c$^2$.
The weak coupling strength of the top quark is not very well constrained,
except that $|V_{tb}|^2\gg |V_{td}|^2+|V_{ts}|^2$ \cite{topR}. Requiring that the
$3\times 3$ CKM matrix is unitary implies that $|V_{tb}|\simeq 1$~\cite{pdg}.
With a matrix of higher rank, though, $|V_{tb}|$ could be small without 
measurably changing the $t\rightarrow Wb$ branching ratio. 
Production of single top quarks provides a direct measurement of $|V_{tb}|$.

\begin{figure}[t]
\begin{center}
  \includegraphics[width=.43\textwidth]{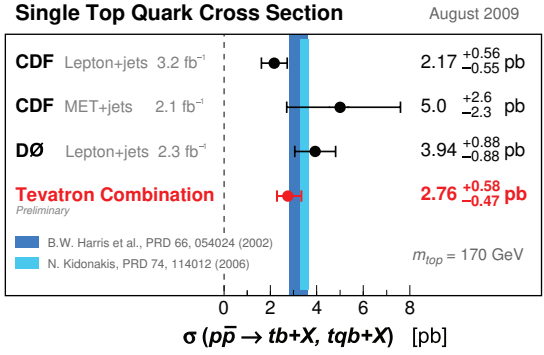}
   \caption{Summary of Tevatron measurements of single top quark production cross section, assuming a top quark mass of 170.0 GeV.}
 \label{tev:cdfstxsec}
\end{center}
\end{figure}

The small cross section of single top quark production and the presence of only one 
top quark in the final state make the separation of the  signal from large 
backgrounds very challenging. Using efficient lepton, jet identification and $b$-quark 
tagging algorithms allows the selection of a candidate sample with a signal to background ratio of about $\sim$~1/20.
\begin{figure}[h]
\begin{center}
  \includegraphics[width=.48\textwidth]{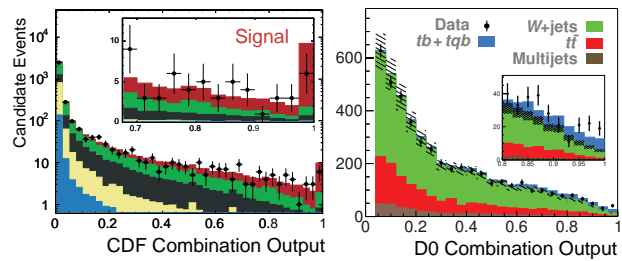}
   \caption{Discriminant distributions for single top quark observation at CDF (left) and D0 (right).
The plot shows the discriminant output of a combination of several input analyses.}
 \label{fig:tevST}
\end{center}
\end{figure}
This poor signal to background ratio renders a simple counting experiment impossible
and demands for better discrimination of signal and background processes.
Consequently, both collaborations have developed sophisticated signal extraction techniques;
a matrix element method, boosted decision trees, neural networks and a likelihood
function approach. The large top quark mass and angular correlations in top quark decay
makes this process very well suited for multivariate analyses.
Careful validation of these techniques in data control samples have been performed.
Multivariate techniques have successfully been used in the single top searches to establish
the first hint of a signal ($\sim$ 2.5 standard deviations) by CDF in October 2006 \cite{dpf2006}, first evidence by D0 in December 2006 \cite{d0evidence} and first evidence by CDF in July 2007 \cite{cdfevidence}. Both collaborations announced jointly the observation of single top
quark production at the 5.0 standard deviations level in March 2009 \cite{stobservation}. Figure \ref{fig:tevST} shows the
discriminant distribution for CDF and D0, obtained by a combination of several input analyses from each experiment
using the same dataset. The analyses assumed a joined signal of $s$-channel and $t$-channel
single top in proportion to the SM prediction. Figure \ref{tev:cdfstxsec}
shows the CDF and D0 single top cross section measurements. A Tevatron combination has been used
to determine the most precise value of the CKM matrix element $V_{tb}$. The cross section obtained in
the Tevatron combination is $\sigma_{s+t}=2.76^{+0.58}_{-0.47}$~pb \cite{stTevComb} consistent with the SM prediction.
The cross section result translates into $V_{tb}=0.91\pm0.08 (stat+syst)$, the most 
precise direct determination of $V_{tb}$ to date.

The two channels of single top production are sensitive to quite
different manifestations of physics beyond the SM. Taken together,
they are a comprehensive probe of the top quark's interactions.
For example the $s$-channel process
is sensitive to models with new heavy particles (e.g. $W'$) leading to resonance
single top production in the $s$-channel while in the $t$-channel a negligible effect is
expected because the additional contributions are suppressed by $1/M_{W^{\prime}}$.
In new physics models with flavor-changing neutral currents,
the $t$-channel rate would be greatly enhanced due to possible ($Ztq/gtq$) couplings
with more favourable access to light flavor parton distribution functions.

\begin{figure}[b!]
\begin{center}
  \includegraphics[width=.34\textwidth]{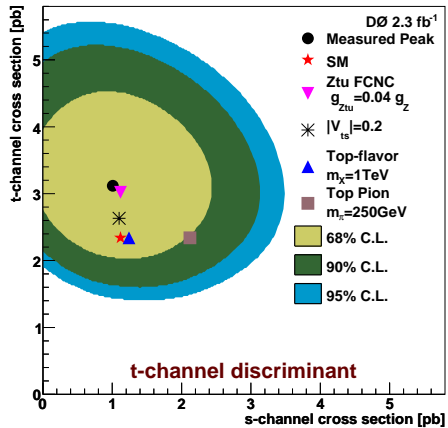}
   \caption{Simultaneous measurement of $s$-channel and $t$-channel single top quark production cross section by D0, assuming a top quark mass of 170.0 GeV/c$^2$ and sensitivity to new physics models.}
 \label{fig:d0tchan}
\end{center}
\end{figure}
The main characteristic of the $t$-channel, which separates it both from the 
$s$-channel process and the backgrounds, is the pseudo-rapidity distribution of the
the light flavor spectator jet, which is very forward. The $s$-channel features a higher fraction of
events where both jets are $b$-tagged compared to the $t$-channel. D0 has recently performed an
updated measurement of the single top cross-section separating $s$-channel
from $t$-channel in a two parameter likelihood fit \cite{d0tchan}.

The posterior probability density is shown in Figure \ref{fig:d0tchan}
Also shown are the SM expectation as well as several representative new 
physics models to illustrate the sensitivity of this analysis \cite{ttait}. 
The signal significance of the $t$-channel in this analysis 
is 4.8 standard deviation, larger than the expected significance of 3.7 standard deviation. 
With larger datasets, this analysis can gain sensitivity to new physics and help identify particular models.

\section{Measurements of the Top Quark Mass} 
The mass of the top quark is a fundamental parameter of the standard model. When combined with the 
$W$~boson mass measurement, it places constraints on the mass of the Higgs boson. This relationship emerges
because quantum loop corrections involving top quarks and the Higgs boson are important for precision 
observables like the mass of the $W$ boson.
\begin{figure}[h!]
\begin{center}
  \includegraphics[width=.35\textwidth]{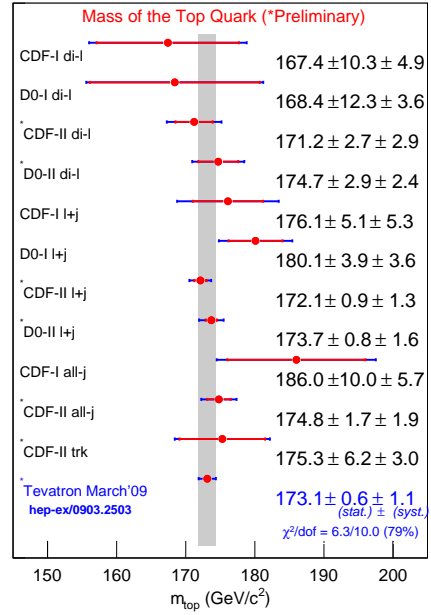}
   \caption{Summary of top quark mass measurements at the Tevatron and their combination.}
 \label{fig:tevmass}
\end{center}
\end{figure}

 Both collaborations have measured the top quark mass in different decay channels and using different 
techniques. Experimental challenges arise from determination of the jet energy scale, the combinatorial 
ambiguity to assign measured jets to the parton of the hard interaction as well as complications due to
QCD radiation.  Both collaborations obtain the most precise results in the lepton + jets channel using the Matrix 
element technique. In this method, an event by event likelihood is calculated based on the differential 
cross section for different top quark mass hypotheses. The combination of all event likelihoods is a very
sensitive estimator of the true top quark mass. The jet energy scale is constrained, {\em in-situ}, through the hadronic decay of 
the $W$ boson present in the top quark decay. A summary of all top quark mass measurements at the Tevatron is 
shown in Figure~ \ref{fig:tevmass}. 
It is worth noting that results among different channels and measurement techniques are consistent.
The Tevatron combined result is $m_t = 173.1 \pm 0.6(stat) \pm 1.1(syst)$ GeV/c$^2$ and has a precision of less than 0.7\% \cite{tevTopMass}. 
There is an ongoing effort at the Tevatron to improve and unfold sources of systematic uncertainty in collaboration with the theory community.

\section{Top Quark Decay} 
\begin{figure}[h]
\begin{minipage}{.45\hsize}
\begin{center}
  \includegraphics[width=.95\textwidth]{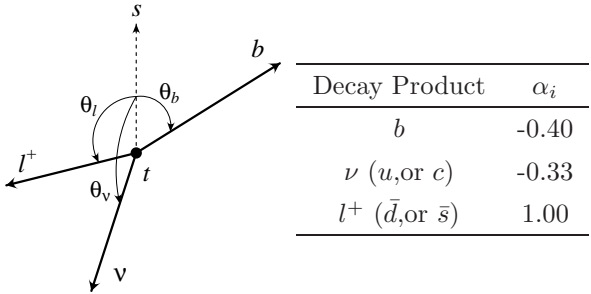}
\end{center}
\end{minipage}
\begin{minipage}{.45\hsize}
\begin{center}
\begin{tabular}{cc} 
\hline
Decay Product & $\alpha_i$  \\   
\hline
$b$ & -0.40\\
$\nu$ ($u$,or $c$) & -0.33\\
$l^{+}$ ($\bar{d}$,or $\bar{s}$) & 1.00\\
\hline
\end{tabular}
\end{center}
\end{minipage}
  \caption{Top quark decay angles in the top quark rest frame and correlation coefficients with the top
quark spin, $\alpha_i$ (table). The direction of the top quark spin is indicated by the vector~$s$. \cite{stspin}}
    \label{fig:topDecayAngle}
\end{figure}

The top quark decays through a charged current weak interaction with a pure left-handed ($V-A$) structure.
Studies of the angular distributions of the top quark decay products provide a direct probe of the nature
of the $Wtb$ vertex. The decay angular distributions are simply linear in the cosine of these decay angles.
\begin{equation}
\frac{1}{\Gamma}\frac{d\Gamma}{d(\cos\theta_i)}=\frac{1}{2}(1+\alpha_i \cos\theta_i)
\label{eqn:pol}
\end{equation}
The degree to which each decay product is correlated with the
spin is encoded in the value of $\alpha_i$ as shown in Figure \ref{fig:topDecayAngle}
The charged lepton is maximally correlated and is the best spin
analyzer of the top quark.

The polarization of the $W$ boson in top quark decays is another sensitive observable.
The longitudinal component is enhanced due to the large Yukawa coupling of the top quark to the 
Goldstone mode of the Higgs field. The fraction of longitudinally polarized $W$ bosons is predicted to be \cite{pdg}
\begin{equation}
f_0 \approx \frac{m_t^2}{2m^2_W+m_t^2}=70.3\pm0.7%
\end{equation}
The fraction of left-handed $W$ bosons is $f_-\approx$ 30\% while the right-handed 
fraction $f_+$ is suppressed by a factor $(m_b/m_t)^2$ and amounts to 
about 3.6$\times$ 10$^{-4}$. Measurements of the $b\rightarrow s\gamma$ decay rate,
assuming the absence of gluonic penguin contributions, have indirectly limited the right-handed 
couplings in top quark decays to less than a few percent \cite{bsg}.

\subsection{Measurement of the $W$ Boson Helicity} 
\begin{figure}[h]
\begin{minipage}{.54\hsize}
\begin{center}
  \includegraphics[width=.99\textwidth]{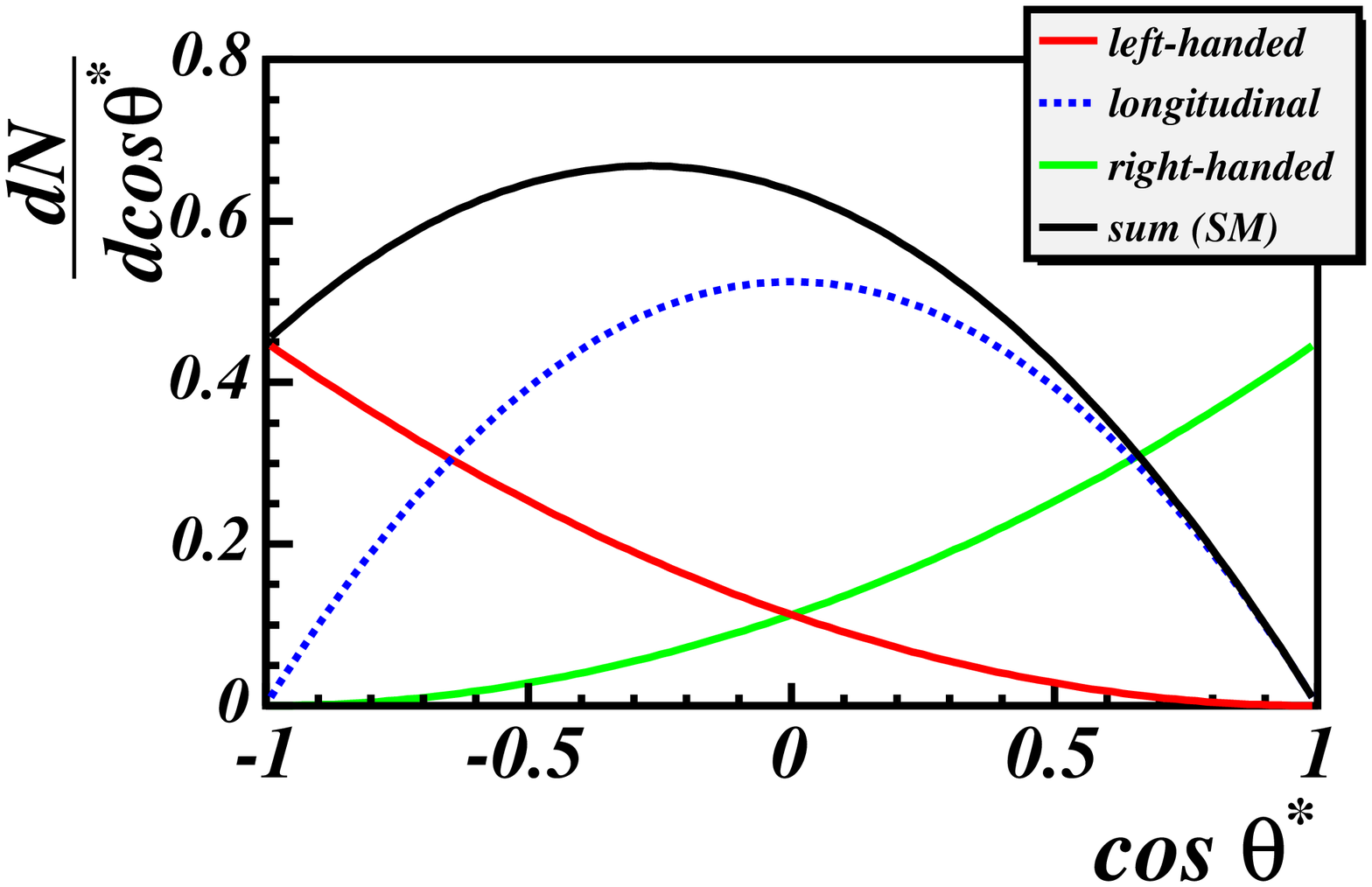}
\end{center}
\end{minipage}
\begin{minipage}{.45\hsize}
\begin{center}
  \includegraphics[width=.99\textwidth]{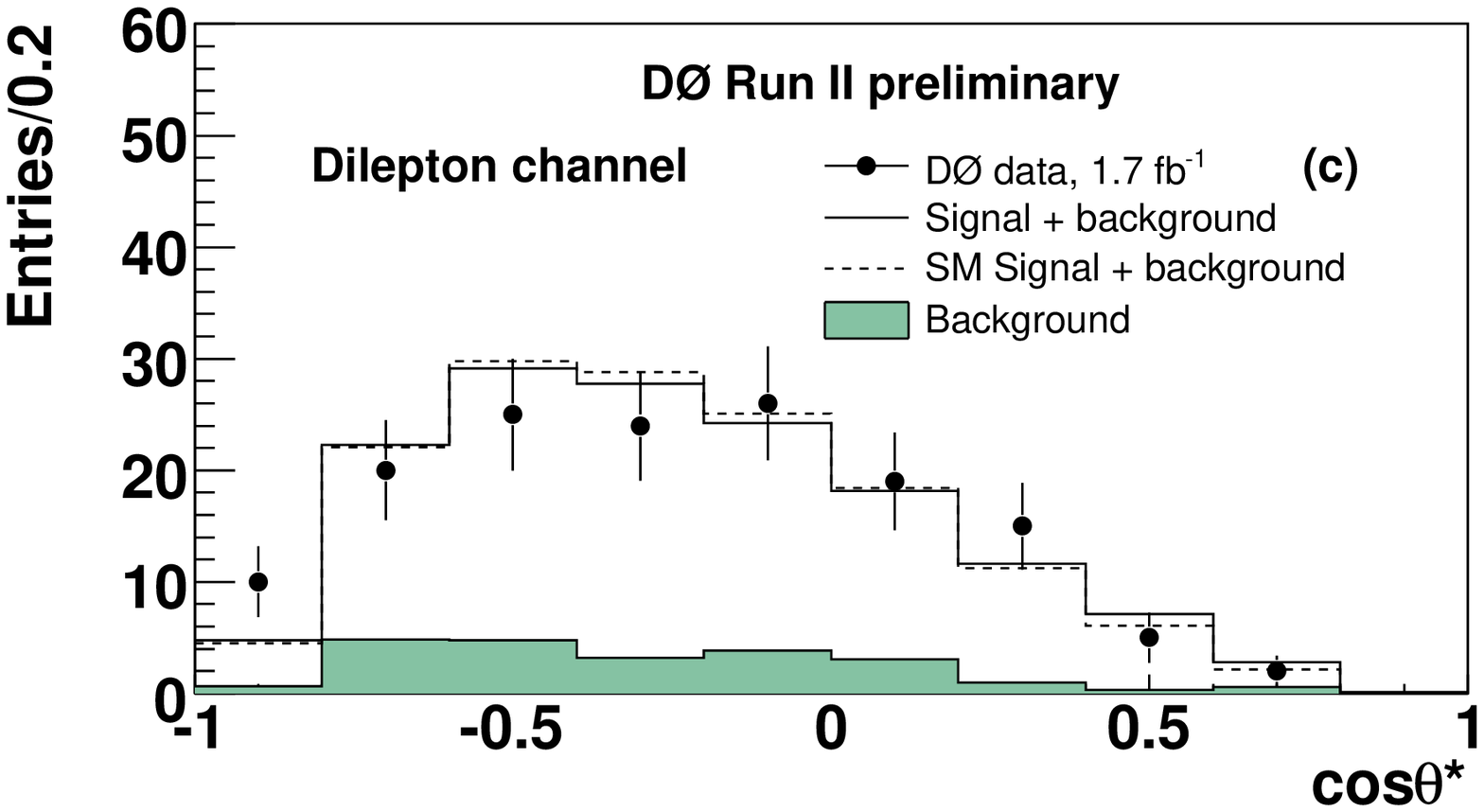}
  \includegraphics[width=.99\textwidth]{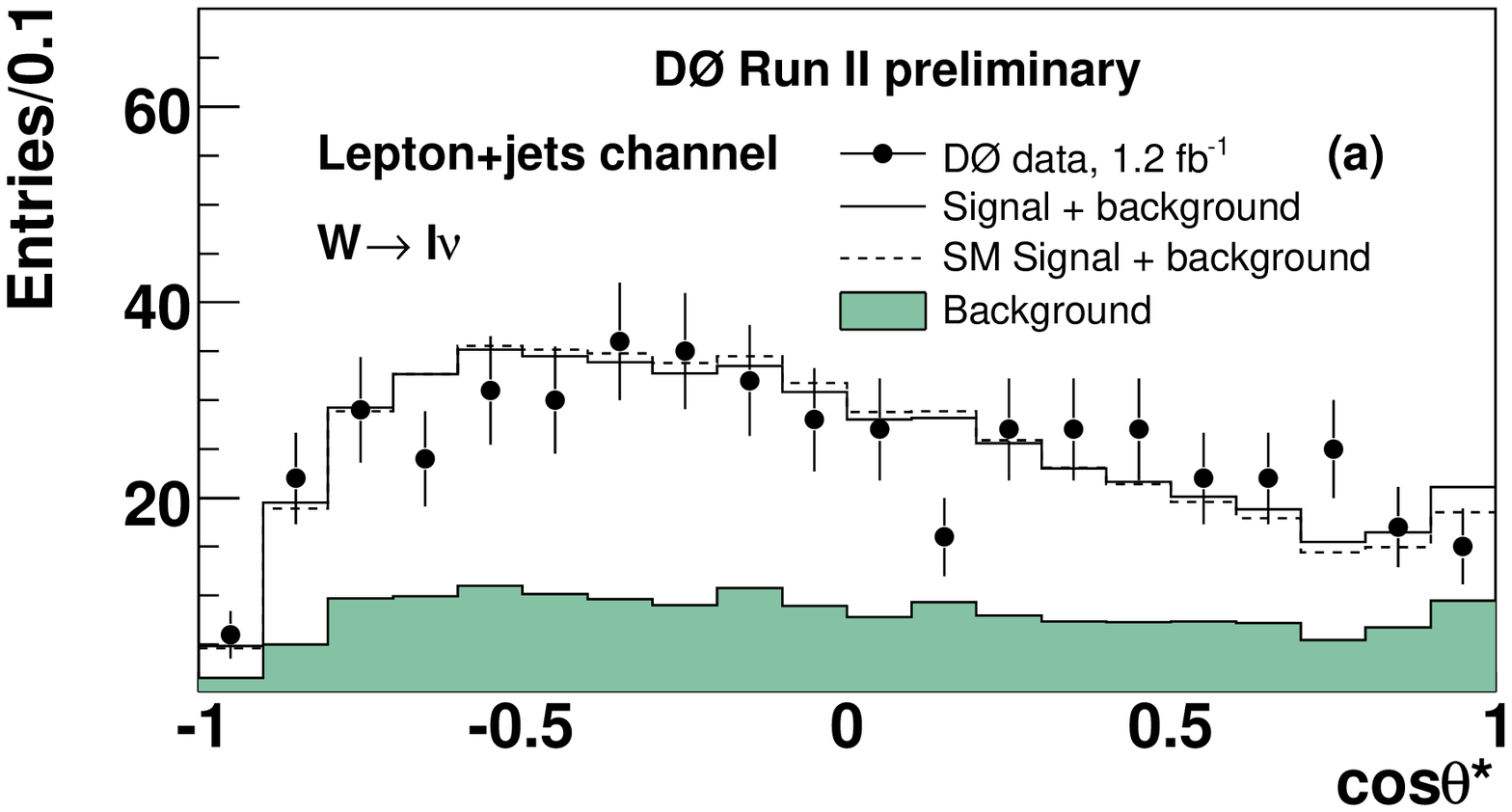}
\end{center}
\end{minipage}
  \caption{Template distributions for $\cos \theta^{*}$ (left) .Comparison of the $\cos \theta^{*}$ distribution in D0 data and the best-fit model for
dilepton events (top right) and lepton + jets events (bottom right). The dashed open histograms show the standard model expectation, and the shaded histograms represent the background contribution.}
 \label{fig:d0Whel}
\end{figure}

The degree of $W$ polarization from top decays can be reconstructed by studying the 
angle $\theta^{*}$ between the charged lepton momentum (defined in the $W$ boson rest frame) and the $W$ 
momentum defined in the top rest frame as shown in Figure \ref{fig:d0Whel}
The dependence of the angular distribution $\omega(\theta^{*})$ on the $W$ boson helicity fractions is given by:
\[
\omega(\theta^{*}) \sim 2(1-\cos^2\theta^{*})f_0 + (1-\cos\theta^{*})^2f_{-}+(1+\cos\theta^{*})^2 f_{+}
\]
CDF and DO have directly measured the $W$ helicity fractions in top quark decay
using $\theta^{*}$. The most model independent technique is to extract the
longitudinal and right-handed fractions simultaneously. D0 measures for the fraction of longitudinal $W$ bosons 
$f_0 = 0.49 \pm 0.11 (stat) \pm 0.09 (syst)$ and for the fraction of right-handed
$W$ bosons $f_+ = 0.11 \pm 0.06 (stat.) \pm 0.05 (syst)$ \cite{d0Whel}. The combination of two similar analyses at CDF results in
$f_0 = 0.66 \pm 0.16 (stat) \pm 0.05 (syst)$ and $f_{+} = -0.03 \pm 0.06 (stat) \pm 0.03 (syst)$ \cite{cdfWhel}.
The results are consistent with the standard model but are still statistically limited.

\section{Top Quark Properties}
\subsection{Top Anti-top Mass Difference}
The CPT theorem is fundamental to any local Lorentz-invariant quantum field theory.
It requires that the mass of a particle and that of its anti-particle be identical.
\begin{figure}[h]
\begin{center}
  \includegraphics[width=.22\textwidth]{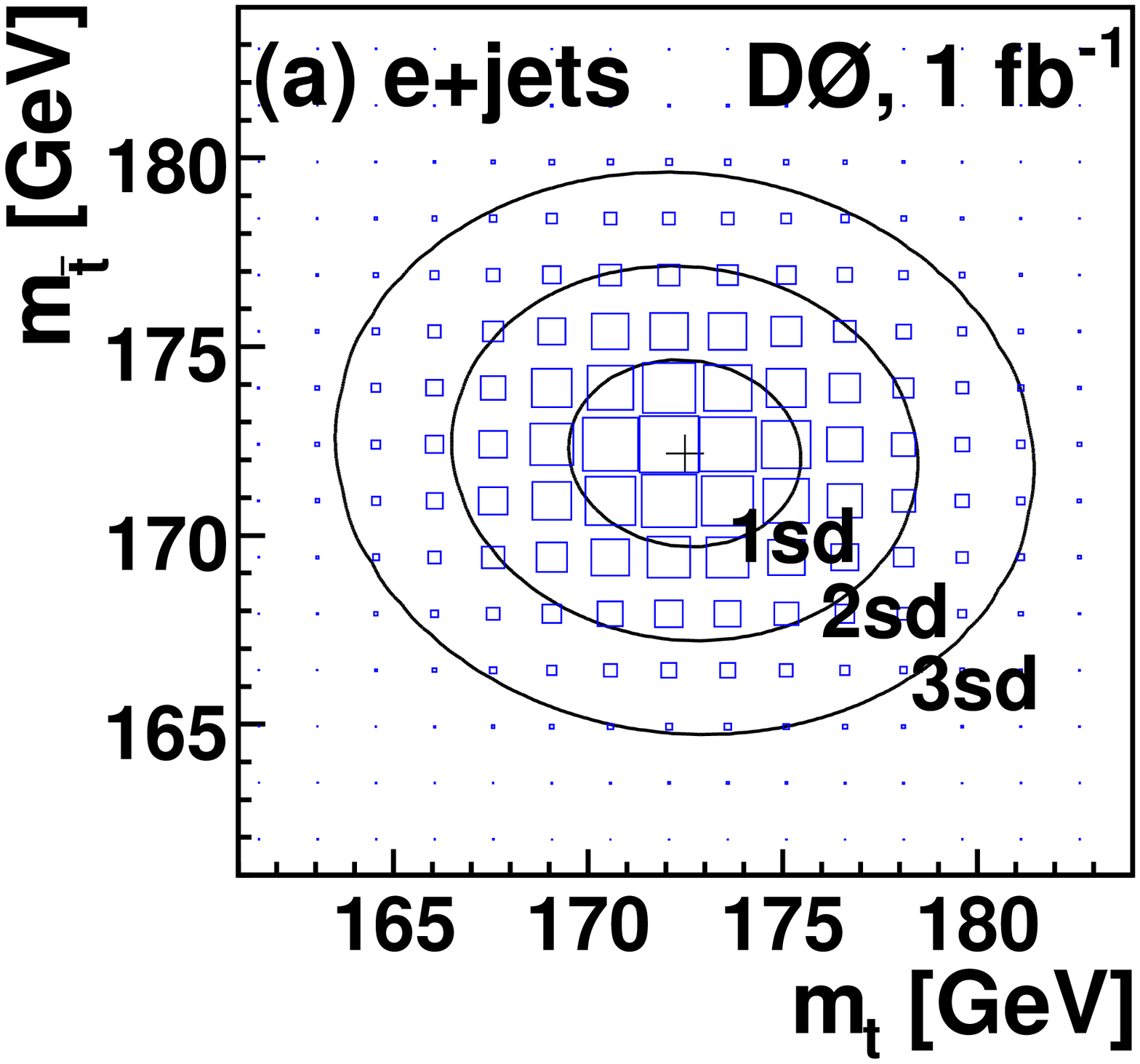}
  \includegraphics[width=.22\textwidth]{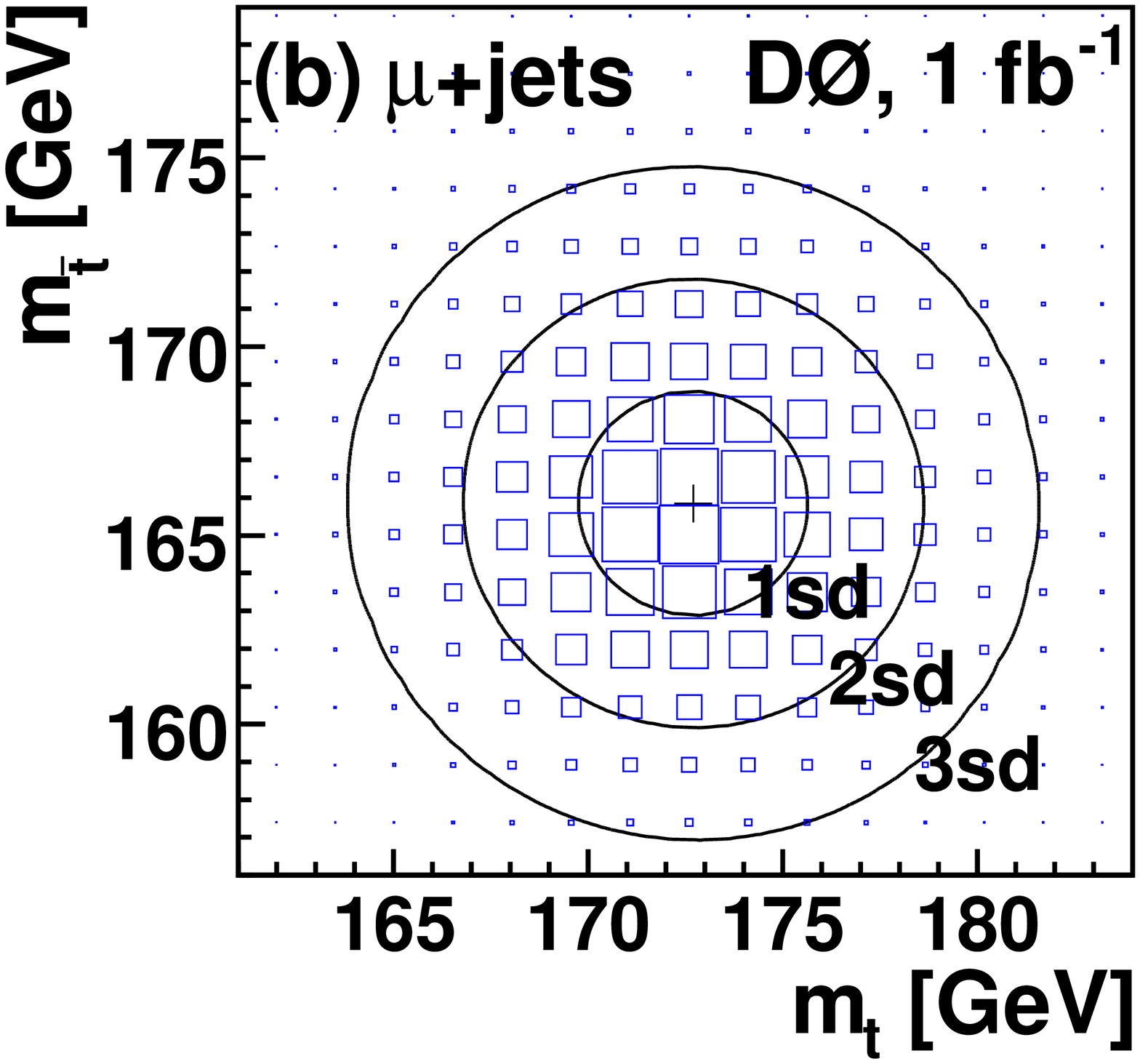}
   \caption{Fitted contours of equal probability for the two dimensional
top quark mass likelihood fit as a function of $m_{\bar{t}}$ and $m_t$ for (a)
e+jets and (b) $\mu$+jets data. }
 \label{fig:d0CPT}
\end{center}
\end{figure}
Tests of CPT invariance for many of the elementary particles accommodated within the standard model
are available in the literature \cite{cpt}. In the quark sector, the mass difference is hard to access because of 
QCD interactions. The short lifetime of the top quark gives us access to a bare quark.
D0 has measured the mass for top and anti-top quarks separately in 1.1 fb$^{-1}$ of data using
lepton + jets data. shown in Figure \ref{fig:d0CPT} Combining the systematic and statistical uncertainties of the measurement
yields to a mass difference of $\Delta = 3.8 \pm 3.7$ GeV/c$^{2}$. The result is consistent with CPT invariance
and represents the first direct measurement of a mass difference between a quark and its anti-quark partner.

\subsection{Spin Correlations in Top Quark Pair Production} 
\begin{figure}[h]
\begin{center}
  \includegraphics[width=.24\textwidth]{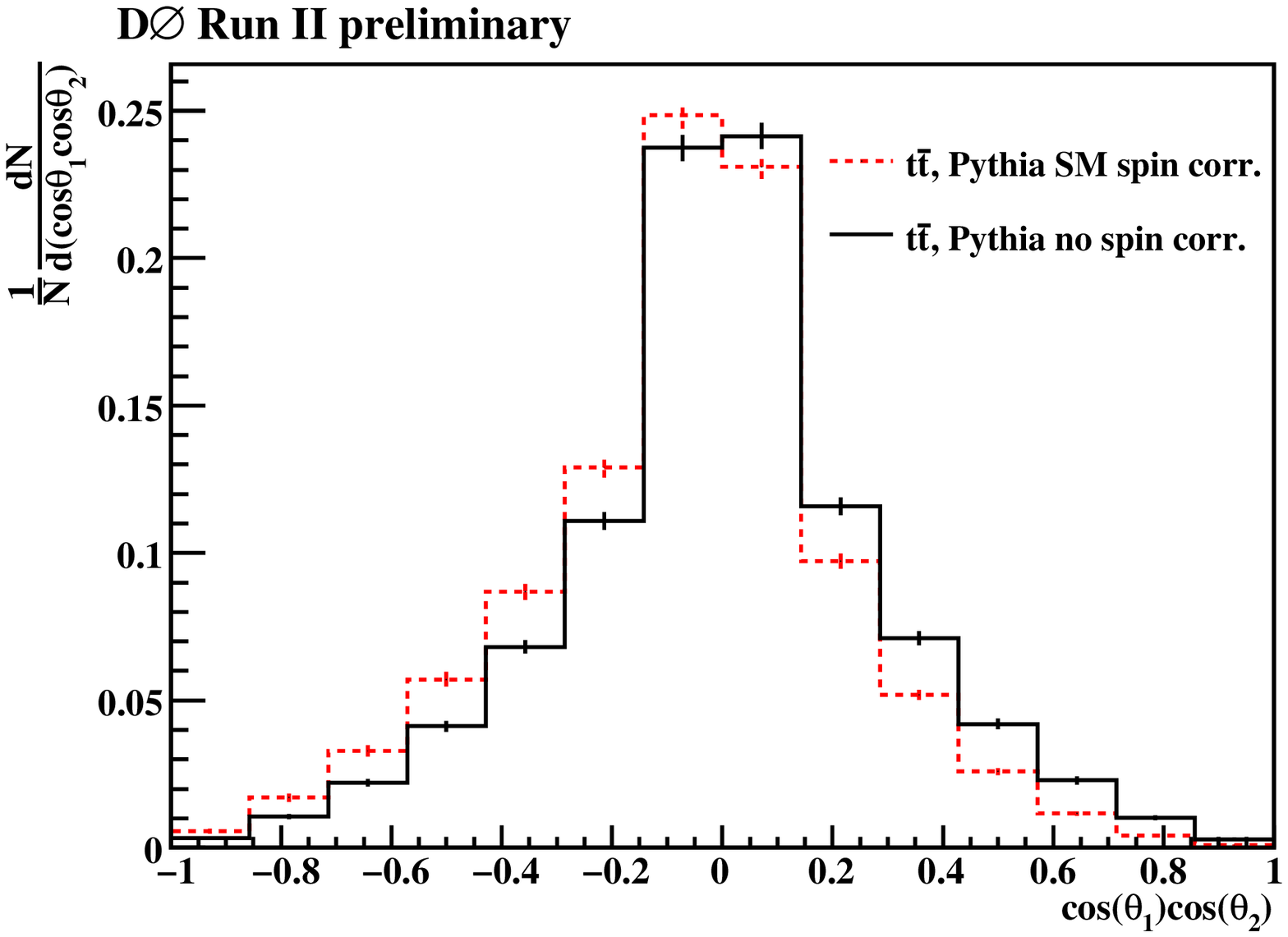}
  \includegraphics[width=.24\textwidth]{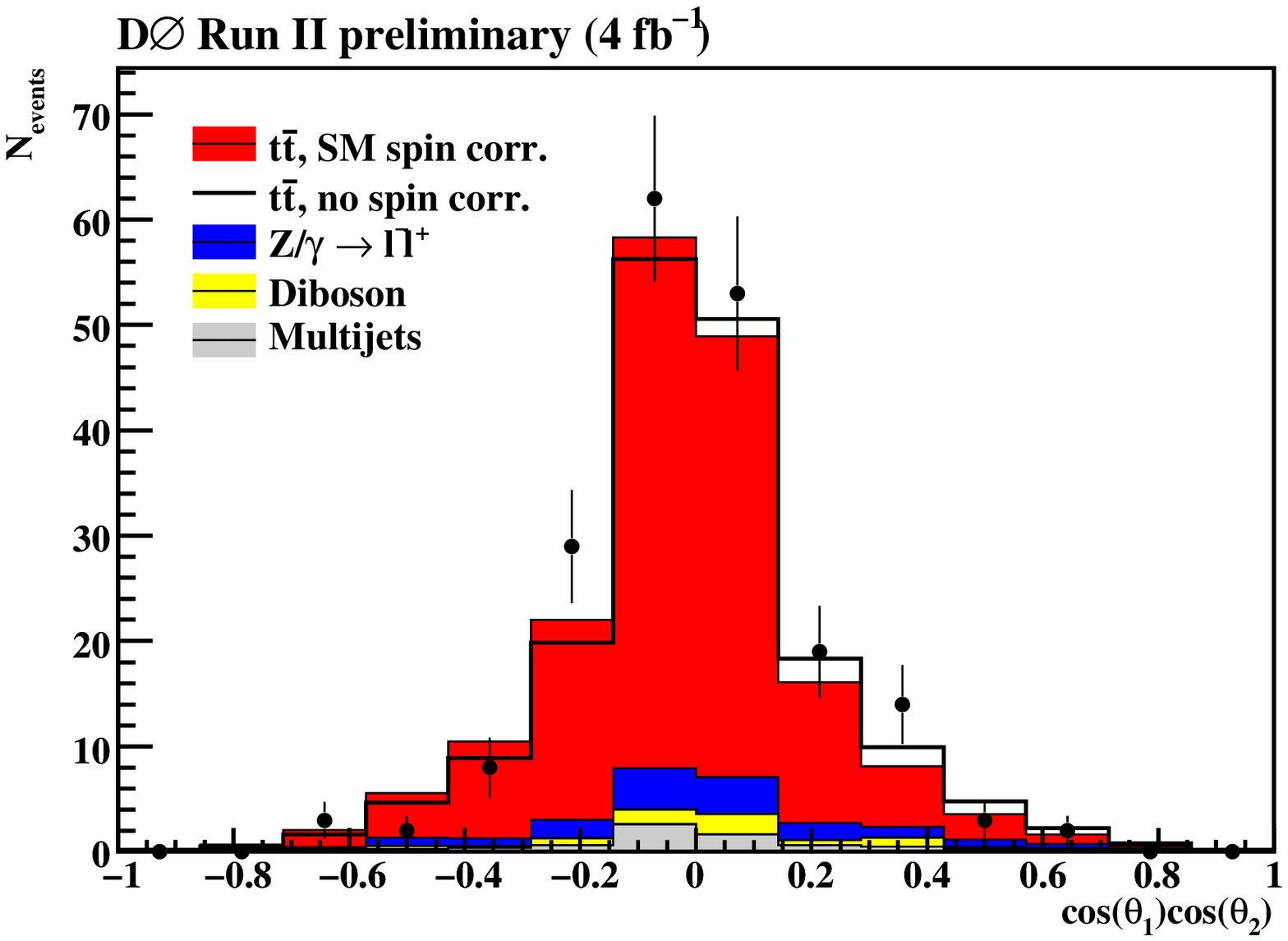}
   \caption{Templates for the double differential distribution $\cos\theta_1 \cos\theta_2$ for dilepton top pair events including NLO QCD spin correlation and without spin correlation (left) and likelihood fit result to the data (right).}
 \label{fig:d0spin}
\end{center}
\end{figure}

One of the most remarkable properties of the top quark is its extremely short lifetime.
Top pair production conserves parity, so the top quark is not polarized 
in its inclusive production, though there are correlations between the top and ¯ 
anti-top quark spin since the production is mediated through a spin-1 gluon.
As shown in Figure \ref{fig:topDecayAngle}, the charged lepton is the best spin analyzer for the top quark.
Spin-spin correlation at $t\bar{t}$ production can therefore be measured through correlations between the flight 
directions of the top quark decay products.  
The correlation coefficient depends on the 
choice of the spin quantization axis. The usual helicity basis measures the component of the top quark spin along
its axis of motion. Because the top mass is large, its helicity
is not a Lorentz invariant quantity. At the Tevatron, the optimal basis was determined to be 
the off-diagonal basis which features a spin correlation coefficient of $\kappa$=0.78 \cite{ofdbasis}. 
The spin correlation coefficient can be extracted fom data, in the dilepton channel, 
by performing a fit to the double differential distribution $\cos\theta_1 \cos\theta_2$ of the
positive and negative lepton direction of flight in the top quark decay as shown in Figure \ref{fig:d0spin}.
Using  2.8 fb$^{-1}$ of data, CDF measures a value of $\kappa = 0.32^{+0.55}_{-0.78}$ assuming 
a top mass of 175 GeV/c$^2$, consistent with the standard model. 
Using the beam axis as the quantization axis, D0 measures $\kappa = -0.17^{+0.64}_{-0.53}$ consistent with the SM at the 2 $\sigma$ level
using 4.0 fb$^{-1}$ of data. These analyses are still statistically limited and would benefit from larger datasets in the near future.   
\subsection{Spin Polarization in Single Top Quark Production} 
\begin{figure}[h]
\begin{center}
  \includegraphics[width=.48\textwidth]{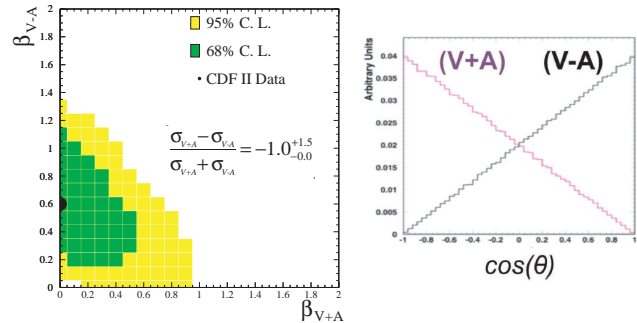}
   \caption{Two dimensional posterior density for left-handed (V-A) vs right-handed (V+A) single top quark
   production (left). Theoretical distribution for $\cos\theta$ in left-handed (V-A) and right-handed (V+A) single top quark production.}
 \label{fig:stpol}
\end{center}
\end{figure}

Single top quark production offers a source of nearly 100\% polarized top quarks
because single top quark production proceeds through a parity violating electroweak vertex
in production and decay. As shown in Figure \ref{fig:topDecayAngle},
the charged lepton (or down type quark) is the best spin analyzer in top quark decay.
In single top events, the angel between the charged lepton in top quark decay and
the down type quark in the initial state follow equation \ref{eqn:pol} \cite{stspin}
when measured in the top quark frame.
This is shown in Figure \ref{fig:stpol} (right) assuming right-handed or left-handed single top quark
production. CDF has recently made a measurement of the top quark polarization using this characteristic.
Because of the low purity in the single top sample, CDF has measured a cross section
assymetry to quantify the single top polarization in production by performing
a cross section fit for candidate events with $\cos\theta>$0 and $\cos\theta<$0.
The two dimensional posterior density for left-handed (V-A) vs right-handed (V+A)
cross section is shown in Figure \ref{fig:stpol} (left). The data favours left-handed
production and the degree of polarization is $pol=\frac{\sigma_{V+A}-\sigma_{V-A}}{\sigma_{V+A}+\sigma_{V-A}}=-1^{+1.5}_{-0.0}$
consistent with the SM prediction, though still statistically limited.

\section{Top Quarks as a Window to New Physics}
\subsection{Resonant Top Quark Pair Production} 
\begin{figure}[h]
\begin{center}
  \includegraphics[width=.45\textwidth]{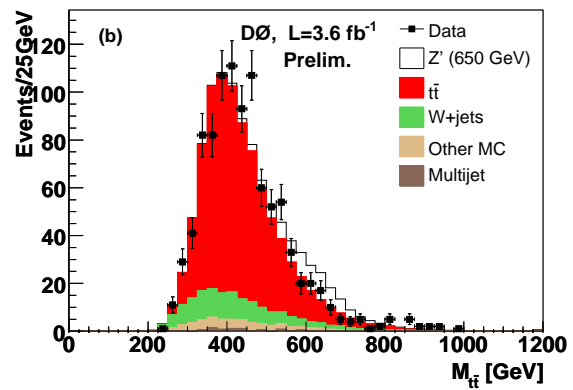}
   \caption{Predicted and observed invariant mass distribution for lepton + $\geq$ 4 jets events with at least
one $b$-tagged jet. Superimposed as white area is the theory signal for a top-color-assisted technicolor $Z^{\prime}$ boson with $M_{Z^{\prime}}$= 650 GeV.}
 \label{fig:d0ttbarRes}
\end{center}
\end{figure}

Several models of new physics predict resonance production of new exotic
particles that decay predominantly into pairs of top quarks, $X\rightarrow t\bar{t}$. 
Such resonant production is possible for massive new particles in extended gauge theories, 
Kaluza Klein states of the gluon or Z boson, axigluons, topcolor \cite{maltoni}.
Independent of the exact model, a narrow width resonance should be visible in the
$t\bar{t}$ invariant mass distribution.
CDF and D0 searched for resonance top quark pair production with a narrow-width 
by investigating the top-pair invariant mass distribution. 
In an analysis using data corresponding to 3.6 fb$^{-1}$ of data, the D0 collaboration 
found no evidence for such a resonance in the lepton + jets channel as shown in Figure \ref{fig:d0ttbarRes}. 
In the absence of any signal D0 places upper limits on $\sigma_X \times BR(X\rightarrow t\bar{t}$)
ranging from 1.0~pb at $M_X$ = 350 GeV/c$^2$ to 0.16 pb at $M_X$ = 1000 GeV/c$^2$. If interpreted in 
the context of a topcolor-assisted technicolor model \cite{topcolor} these limits can be used to derive mass 
limits on a narrow lepto-phobic $Z^{\prime}$ with $M_{Z^{\prime}} >$ 820 GeV/c$^2$ at the 95\% C.L., assuming $\Gamma(Z^{\prime})$ = 0.012 $M_{Z^{\prime}}$.
A similar analysis in the all-hadronic channel at CDF yields similar mass limits.
\subsection{Forward Backward Asymmetry} 
Quantum chromodynamics at leading order predicts that 
the top quark production angle is symmetric with respect to beam direction. 
At NLO, QCD predicts a small charge asymmetry, $A_{fb} = 0.050 \pm 0.015$ \cite{AfbTheory}, 
due to interference of initial and final-state radiation
diagrams and interference of box diagrams with the Born level process. 
In the CP invariant Tevatron frame, the charge
asymmetry is equivalent to a forward-backward asymmetry of the 
produced top quarks. CDF has recently updated a measurement of
the forward backward asymmetry using 3.2 fb$^{-1}$ of data.
The asymmetry is measured through the rapidity $y_{had}$ of 
the hadronically-decaying top (or anti-top) quark and by tagging 
the charge of the top quarks through the lepton charge $Q_l$
as shown in Equation \ref{eqn:afb}.
\begin{equation}
A_{fb}=\frac{N(-Q_l\cdot y_{had}>0)-N(-Q_l\cdot y_{had}<0)}{N(-Q_l\cdot y_{had}>0)+N(-Q_l\cdot y_{had}<0)}
\label{eqn:afb}
\end{equation}
\begin{figure}[b!]
\begin{center}
  \includegraphics[width=.42\textwidth]{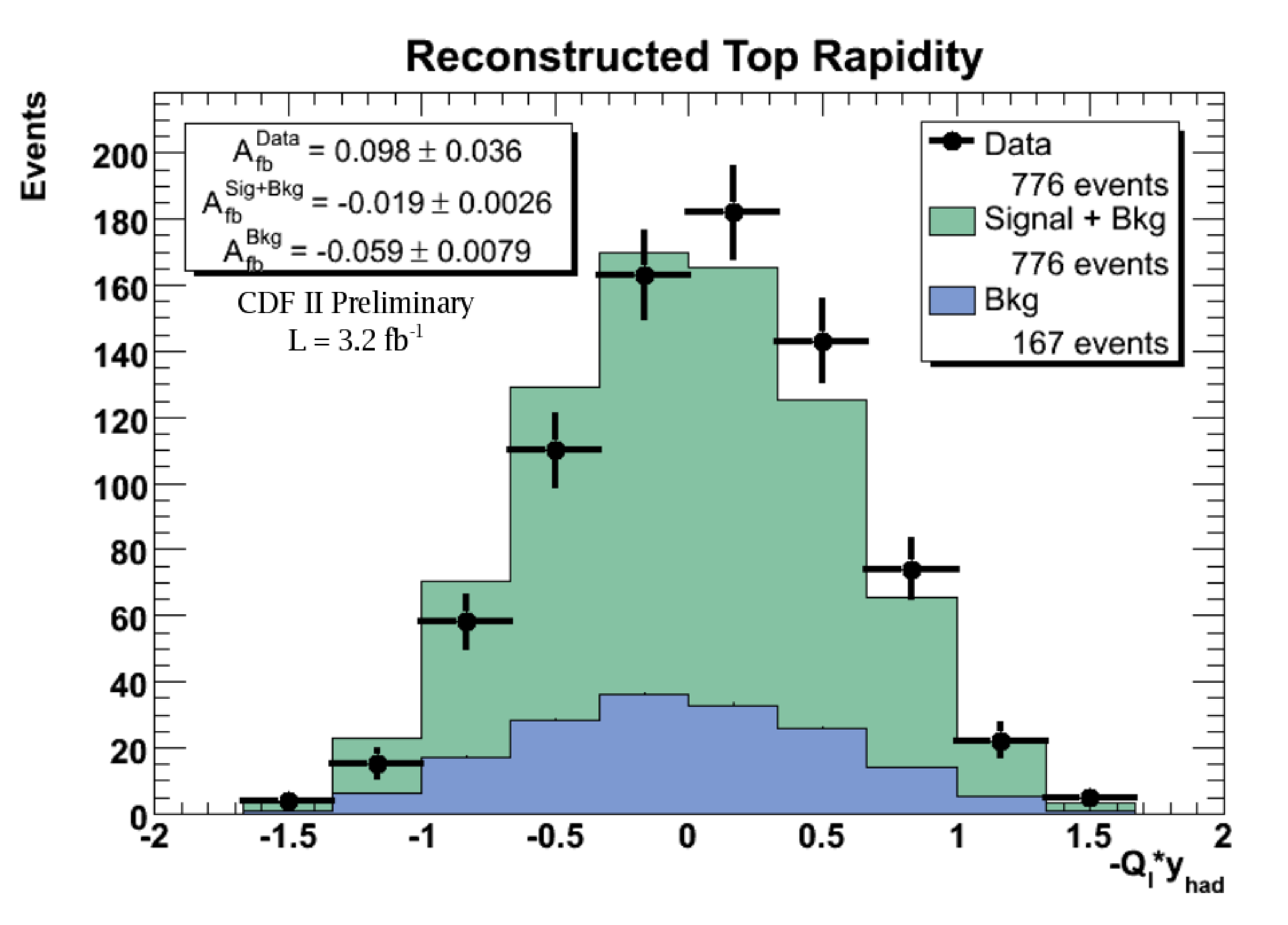}
   \caption{Distribution of $-Q_l\cdot y_{had}$ for data and prediction.}
 \label{fig:afbCDF}
\end{center}
\end{figure}
CDF measures a forward-backward asymmetry 
in the proton anti-proton lab frame of $A_{fb}^{det} = 0.193 \pm 0.065 (stat) \pm 0.024 (syst)$.

The relatively large value compared to the SM expectation confirms earlier CDF 
measurements \cite{cdfAfb}. The CDF measurement quoted above is corrected for background contributions, 
acceptance and reconstruction biases. D0 has also measured an asymmetry. 
The result obtained after applying a background correction, is $A_{fb} = 0.12 \pm 0.08 (stat) \pm 0.01(syst)$. 

The Tevatron results indicate some tension with the standard model prediction. New
physics could give rise to an increased asymmetry through interference of new particles
at high energy not yet directly observed. More data will be required to differentiate 
statistical fluctuations from new physics.

\subsection{Search for Heavy t$^{\prime}$ Quarks} 
\begin{figure}[h]
\begin{center}
  \includegraphics[width=.42\textwidth]{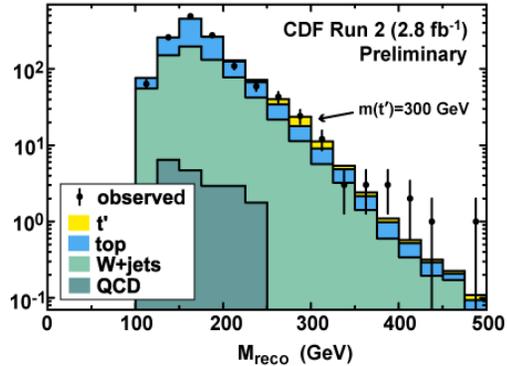}
   \caption{Distribution of the reconstructed mass of the heavy quark for a hypothetical $t^{'}$  signal and background. The $t^{'}$ is normalized to the 95\% CL upper limit.}
 \label{fig:tprime}
\end{center}
\end{figure}

Using 2.8 fb$^{-1}$ of integrated luminosity CDF searched for pair production of new 
heavy top-like quarks ($t^{\prime}$). A forth chiral generation of massive fermions is predicted in 
a number of models, such as two-Higgs-doublet scenarios and N=2 SUSY models \cite{fourthGen}. 
In this search, a small mass splitting between new heavy $t^{\prime}$ and $b^{\prime}$ quarks is preferred, such that
the $t^{\prime}$ decays predominantly to a $W$ boson and a down-type quark ($q=d,s,b$). Such heavy exotic quarks 
are present in other models as well, like the beautiful mirrors \cite{bMirror}, and 
little Higgs models with conserved T-parity \cite{littleH}. Assuming this new quark is pair-produced 
strongly, has a mass greater than the top quark and decays predominantly to $Wq$, its 
event signature would mimic that of the top pair production. To distinguish the $t^{\prime}$ signal 
from the SM backgrounds, a 2D fit to the $t^{\prime}$ reconstructed mass, as shown in Figure~\ref{fig:tprime}
and the scalar sum of the total transverse energy is performed. No $t^{\prime}$ signal is observed above
background predictions and 95 \% C. L. upper limits on the $t^{\prime}$ production rate are set, excluding a $t^{\prime}$ quark with a mass below 
311 Gev/c$^2$. 
\subsection{Search for sTop sQuarks in the Top Quark Sample}
Supersymmetry (SUSY) is one of the most plausible extensions to the SM
because it naturally solves the hierarchy problem and provides a natural dark matter candidate.
To reconcile SUSY with experimental data, SUSY must be broken, and
the SM particles must obtain their super-partners with distinct, mostly much heavier, masses.
However, due to the large mass of the top quark, the mass splitting between SUSY partners
can be large, such that the lighter stop squark is in fact the lightest squark of all, and can
possibly be even lighter than the top quark. Since stop squarks are scalar particles, the pair 
production cross-section is about an order of magnitude smaller than for a fermionic quark of similar mass.
\begin{figure}[h]
\begin{center}
  \includegraphics[width=.40\textwidth]{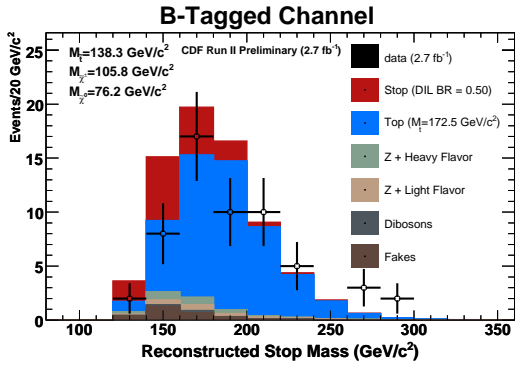}
  \caption{Reconstructed stop mass for stop signal (normalized to the dilepton branching ratio excluded at the 95\% level)
        for various chargino, neutralino, and stop masses. The data is consistent with the standard model.}
    \label{fig:stopReco}
\end{center}
\end{figure}
CDF has performed a search for stop quarks with the possibility that $m_{\tilde{t}}\leq m_t$,
favored by super-symmetric electroweak baryogenesis scenarios~\cite{barSUSY}, and that the 
lightest supersymmetric particle (LSP) is the neutralino, as favored by the astrophysical data \cite{pdg}. 
If the chargino is lighter than the stop quark with $m_{\tilde{\chi}^{\pm}} < m_{\tilde{t}} - m_b$, the decay
$\tilde{t}\rightarrow b\tilde{\chi}^{\pm}_1$ dominates. This results in a topology of stop quarks that will decay into a final state
similar to that of pair produced top quarks. In the dilepton channel, two additional neutral
particles, the LSP, will be present  $\tilde{t}\rightarrow b\tilde{\chi}^{\pm}_1\rightarrow b\tilde{\chi}^{0}_1 l\nu$.
The mass splitting between the chargino and the neutralino modifies the lepton branching ratio, which can
significantly be enhanced over the standard model prediction for top quark decay. The search is performed by
looking for an excess in the reconstructed stop mass distribution as shown in Figure \ref{fig:stopReco}
\begin{figure}[h]
\begin{center}
     \includegraphics[width=.24\textwidth]{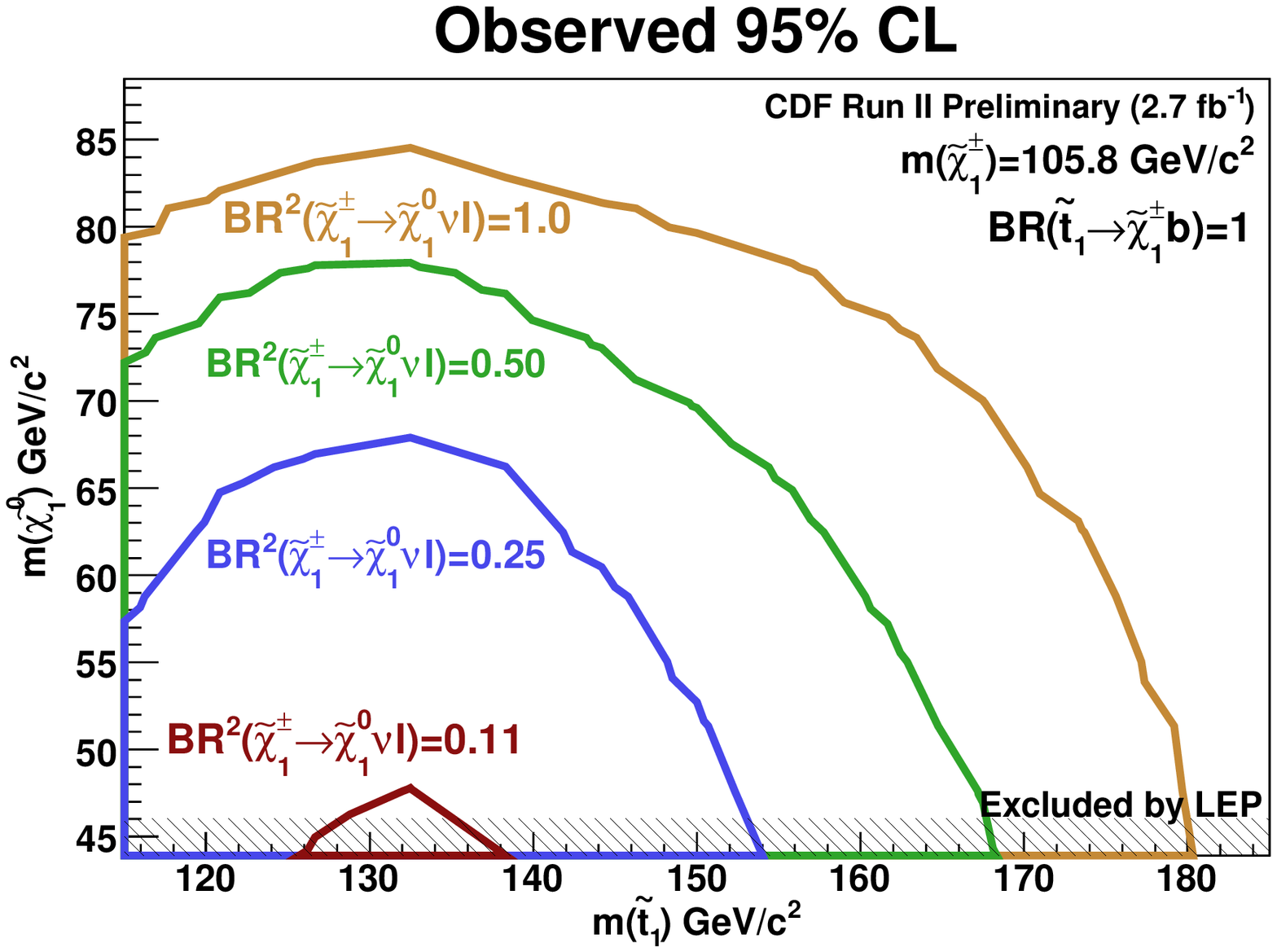}
     \includegraphics[width=.24\textwidth]{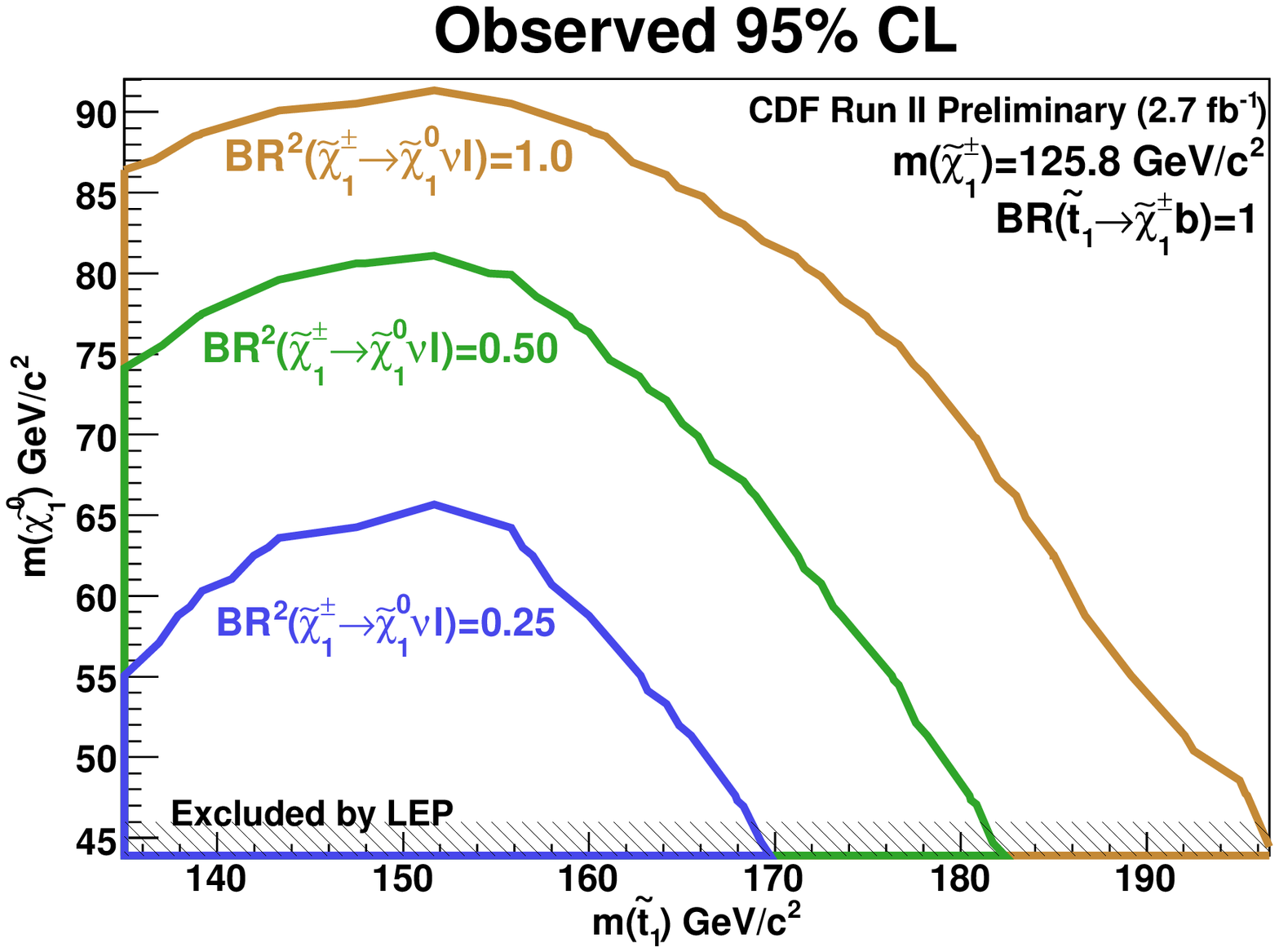}
      \caption{Observed 95\% CL in the Neutralino Mass v. Stop Mass plane, for various dilepton branching ratios
        at chargino masses of 105.8 GeV (left) and 125.8 GeV (right).}
    \label{fig:stopLimit}
\end{center}
\end{figure}
Figure \ref{fig:stopLimit} shows the observed 95\% C. L. limit in the neutralino-mass versus 
stop-mass plane, for various dilepton branching ratios.

\subsection{Search for Charged Higgs in Top Quark Decays}
In many extensions of the standard model, including SUSY and grand unified theories,
the existence of an additional Higgs doublet is required. Such models predict multiple physical Higgs
particles, including three neutral and two charged Higgs bosons \cite{chargedH}. If a charged Higgs boson is sufficiently
light, it can appear in top quark decays $t\rightarrow H^{+}b$.
CDF has performed an analysis based on the kinematics of events in the lepton plus
jets decay mode with two $b$-tagged jets. This search is sensitive to the decay $H^{+}\rightarrow c\bar{s}$, 
which is dominant in the minimal supersymmetric standard model (MSSM) for small values of
$tan \beta$. CDF searches for an $H^{+}$ resonance in the invariant mass of the two leading non-tagged jets using 
2.2 fb$^{−1}$ of data. In the absence of any excess, upper limits are set
on $B(t\rightarrow H^+b)$ at 95\% C.L. of 0.1 to 0.3 assuming $B(H^+\rightarrow c\bar{s})$ = 1.0 for charged Higgs
masses of 60 to 150 GeV/c$^2$ \cite{cdfChargedHiggs}. 
\begin{figure}[h]
\begin{center}
  \includegraphics[width=.24\textwidth]{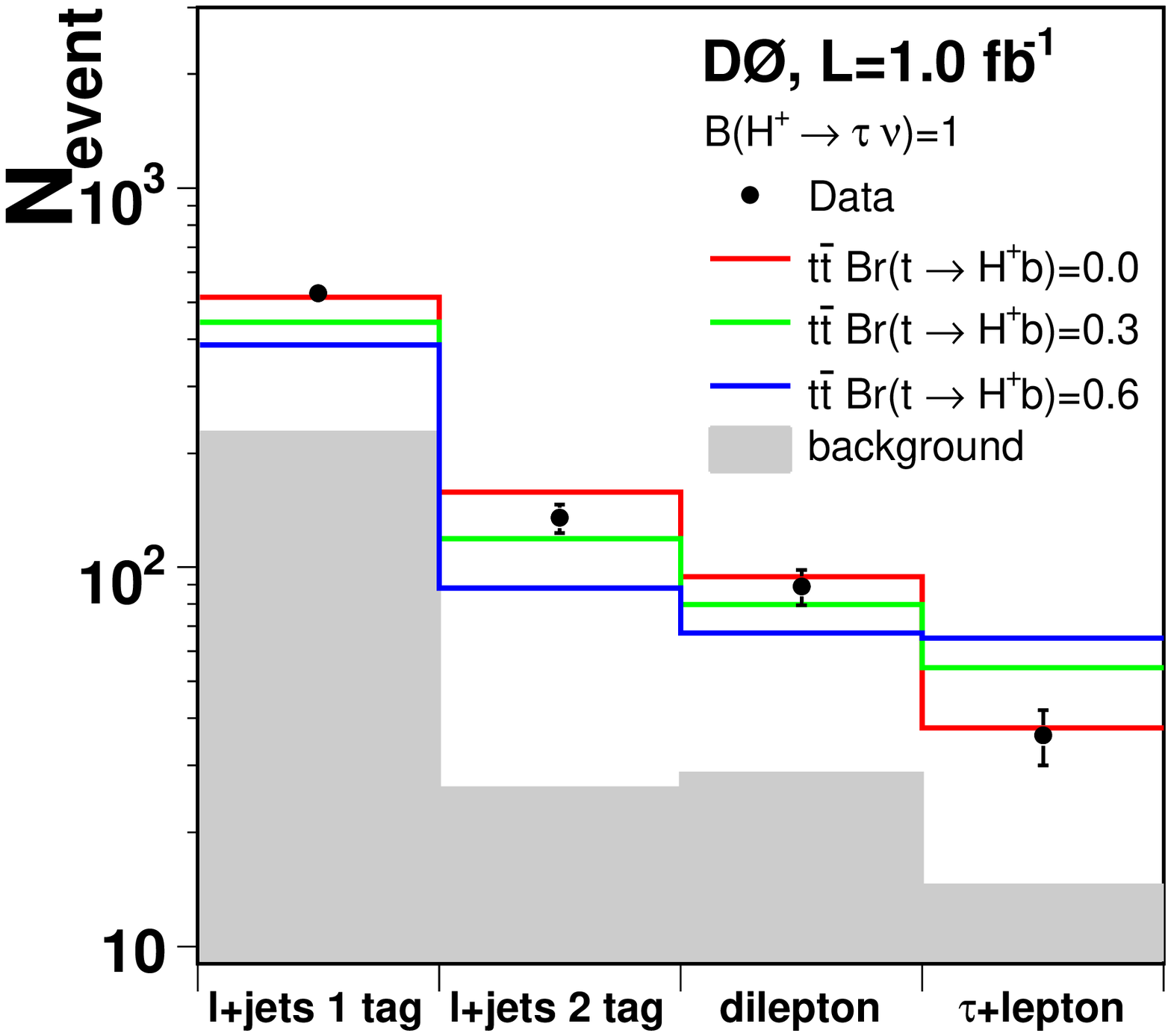}
  \includegraphics[width=.24\textwidth]{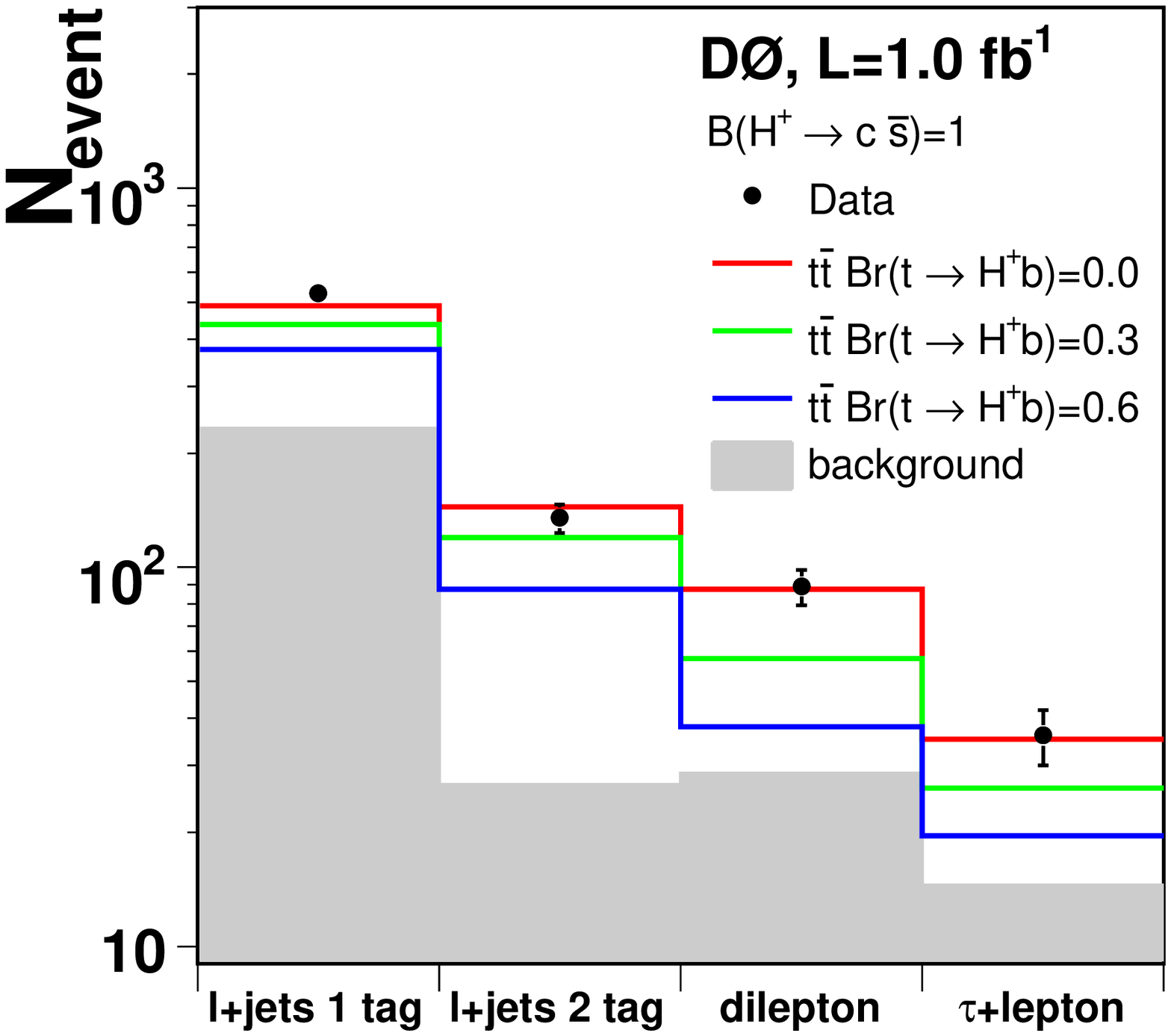}
   \caption{Number of expected and observed events versus final
state signature for $M_{H+}$ = 80 GeV assuming either exclusive $H^+\rightarrow \tau^{+}\nu$  (left) 
or exclusive $H^+\rightarrow c\bar{s}$  of the charged Higgs boson.}
 \label{fig:d0chargedHiggs}
\end{center}
\end{figure}

D0 utilizes an alternative analysis strategy. Using 1.0~fb$^{-1}$ of data, a global analysis
is performed using the event yields in different top quark final states. Event yields in the
$e$+jets, $\mu$+jets, $ee$, $e\mu$, $\mu\mu$, $e\tau$ and $\mu \tau$ channels are 
compared to the expectation from the SM as shown in Figure \ref{fig:d0chargedHiggs}. The yields 
are sensitive to contributions of charged Higgs through $H^{+}\rightarrow \tau^{+}\nu$ and $H^{+}\rightarrow c\bar{s}$
which dominates for large and small values of $\tan \beta$ in the MSSM, respectively. 
D0 observes no significant deviation from expectations and sets upper limits 
at 95\% C. L. on the branching ratios $B(t\rightarrow H^+b)$ for different values of
$B(H^+\rightarrow \tau^{+}\nu)$ and $B(H^+\rightarrow c\bar{s})$.
In the lepto-phobic scenario, branching ratios $B(t\rightarrow H^+b)>$ 0.22
are excluded for the $M_{H^{+}}$ range between 80 and 155 GeV.
\begin{figure}[b]
\begin{center}
  \includegraphics[width=.4\textwidth]{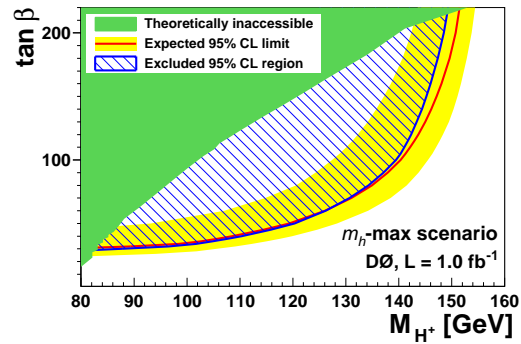}
  \caption{Excluded region in the $\tan\beta, M_{H^{+}}$ parameter space in
the MSSM for the $mh-max$ scenario. The yellow band shows
the $\pm$ 1 SD band around the expected limit.}
 \label{fig:modelChargedHD0}
\end{center}
\end{figure}
In this scenario, D0 has
also performed a model-independent measurement and
excluded $B(t\rightarrow H^+b) >$ 0.12 - 0.26 depending on $M_{H^{+}}$.
The results are interpreted in different models and limits are set
in [$\tan\beta, M_{H+}$] parameter space. For example in the
$m_h-max$ scenario, $M_{H^{+}}$ up to 149 GeV/c$^2$ are
excluded as shown in Figure \ref{fig:modelChargedHD0} These are the most restrictive limits to date
in direct searches for charged Higgs bosons in top quark decays.

\section{Top Quark Physics at the LHC}
Compared to the Tevatron, the top quark production cross-sections at the LHC are
expected to be two orders of magnitude larger, making the LHC a top quark factory. 
At 14 TeV, the production cross section is expected to be about 830~pb for top quark pair production 
and 315~pb for single top quark production. If the LHC is operating at 10 TeV during
the first year, the top quark production cross sections is expected to be roughly half.

Both LHC experiments, ATLAS and CMS, have prepared multiple top physics analyses and tested them on simulated 
data in anticipation of the collider turn on \cite{topCMS,topATLAS}. 
In the absence of reliable $b$-quark tagging in early data, the dilepton
channel is likely to be the top quark re-discovery channel with first LHC data.

\begin{figure}[h]
\begin{center}
  \includegraphics[width=.33\textwidth]{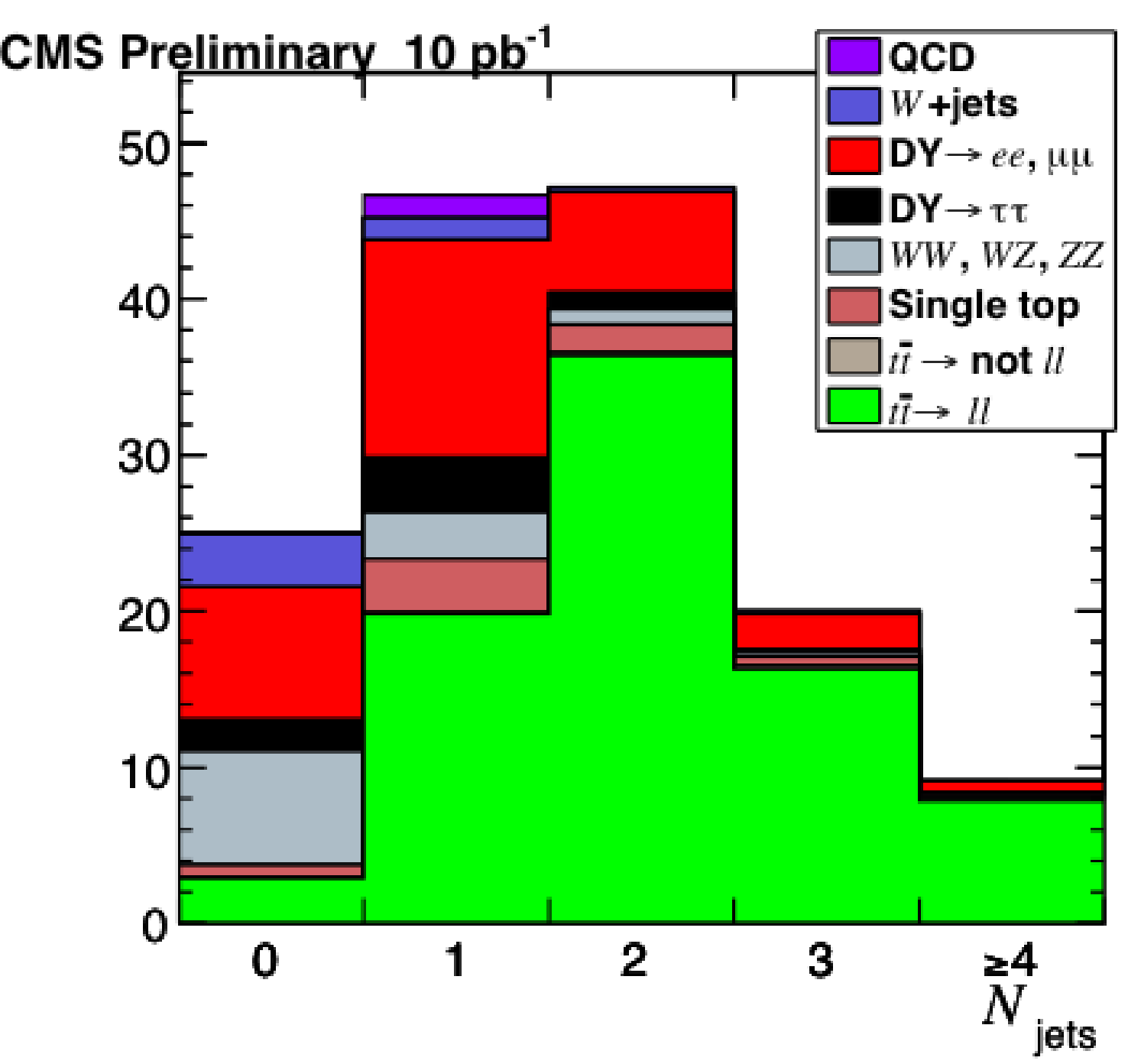}
  \caption{Expected number of dilepton events at CMS as a function of jet multiplicity
based on MC simulation and normalized to 10 pb$^{-1}$ of data in 10 TeV LHC collisions.}
 \label{fig:cmsDIL}
\end{center}
\end{figure}

\begin{figure}[b]
\begin{center}
  \includegraphics[width=.36\textwidth]{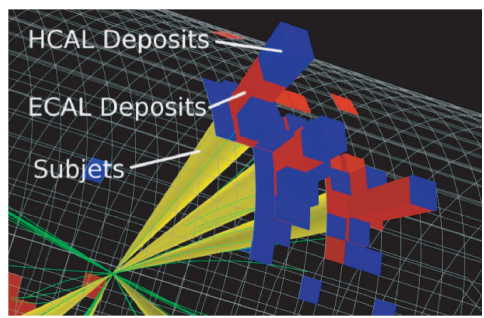}
  \caption{Reconstructed top-quark jet in cylindrical view with $p_T$ = 800 GeV/c. The cones
represent the subjets. The calorimeter deposition, and the subjets are indicated on the
figure.}
 \label{fig:cmsLJ}
\end{center}
\end{figure}

CMS has recently updated their expectations for observing top quark pair production in final states with two leptons and jets using the first 10 pb$^{-1}$  of CMS data \cite{topCMS} as shown in Figure \ref{fig:cmsDIL} Clear observation of top quark production is expected
in less than 10 pb$^{-1}$ of data at 10 TeV LHC collisions. A signal-to-noise ratio of about 4 to 1 is expected in
all channels combined and about 9 to 1 in the electron-muon channel alone. 
The signal production cross section is expected to be measured with an uncertainty of $15\%(stat) \pm 10\%(syst) \pm 10\%(lumi)$ 
in this dataset.
The high center of mass energy at the LHC will considerably extend the reach to discover new heavy particles over the Tevatron.
If these new particles are sufficiently massive, the resulting top quarks are highly boosted, and may 
collapse into a single top quark jet as shown in Figure \ref{fig:cmsLJ}
It is therefore useful to develop reconstruction algorithms that attempt to distinguish these boosted top quarks from
those produced in generic QCD background.

The general strategy for tagging boosted top quarks decaying hadronically is to identify jet
substructure in top-quark jets and decompose them into subjets. The substructure information
can be used to impose kinematic cuts that discriminate against non-top jets of the same $p_T$. 
In particular, the reconstructed $W$ boson and top quark mass provide powerful discrimination.

The CMS collaboration has developed an algorithm that makes boosted top quarks with hadronic decay 
accessible experimentally, due to its high rejection
($\sim$ 98\%) of background jets with $p_T$ = 600 GeV/c while retaining a high fraction
of boosted top-quark jets ($\sim$ 46\%) with the same $p_T$) \cite{topCMS}. This performance is comparable
to that for $b$-quark tagging algorithms at hadron colliders.
\section{Conclusions}
The discovery of the top quark opened up a rich field in particle physics.
A broad top physics program is underway at the Tevatron \cite{topTevatron}.
So far, the data seems consistent with the standard model.
While the precision of the top quark pair production cross section
has surpased the theoretical prediction, several top quark properties measurements
are still statistically limited. 
The Tevatron expects to double the available data sets if the experiments
are running through 2011.
With the observation of electroweak single top quarks a new source of
top quarks is now available at the Tevatron. Top quark property measurements
using the single top sample are interesting and have already lead to the most precise direct determination of $V_{tb}$.
The precision on the top quark mass has reached 0.75\%.
With the LHC, an enormous top quark factory is beginning operation, prepared to
continue to explore the truth about the top quark.

\end{document}